\documentclass{ws-ijmpb}

\begin{document}

\markboth{B. K. Raj, B. Pradhan and G. C. Rout}
{Dynamic Jahn-Teller effect and superconducting......}

%%%%%%%%%%%%%%%%%%%%% Publisher's Area please ignore %%%%%%%%%%%%%%%
%
\catchline{}{}{}{}{}
%
%%%%%%%%%%%%%%%%%%%%%%%%%%%%%%%%%%%%%%%%%%%%%%%%%%%%%%%%%%%%%%%%%%%%

\title{DYNAMIC JAHN-TELLER EFFECT AND SUPERCONDUCTING EXCITATIONS IN RAMAN SPECTRA}

\author{B. K. RAJ}

\address{Department of Physics, Government (Autonomous) College,\\
 Angul-759128, India}

\author{B. PRADHAN\footnote{Corresponding author, Email: brunda@iopb.res.in, Mob: +91-9437806565}}

\address{Department of Physics, Government Science College,\\
 Malkangiri-764048, India}

\author{G. C. ROUT\footnote{Email: gcr@iopb.res.in: Mobile: +91-9937981694}}

\address{Department of  Physics, School of Applied Sciences,\\
 KIIT, University, Bhubaneswar-751024, India.}
\maketitle

\begin{history}
%\received{xxx}
%\revised{xxx}
%\accepted{(Day Month Year)}
%\comby{(xxxxxxxxxx)}
\end{history}

\begin{abstract}
We report a model study of the phonon response on the interplay of
 the superconducting (SC) gap and both static and dynamic Jahn-Teller
 distortion and predict the appearance of the SC and Jahn-Teller (JT)
 distortion gap excitation peaks in the Raman spectra of high-T$_c$
 superconductors. The model consists of the Hamiltonian containing
 static JT interaction and $s$-wave type SC interaction in the 
conduction band. Further the phonons are coupled to the density 
of the conduction electron in the band as well as to the JT split 
conduction band giving rise to dynamic Jahn-Teller (DJT) interaction.
 The phonons are considered in a harmonic approximation. The phonon 
Green's function is calculated by Zubarev's technique and the phonon
 self-energy arising due to normal electron-phonon (EP) interaction 
and DJT interaction. The phonon self-energy arising due to this is 
calculated from the electron response density function in the coexistence
 phase of the two order parameters. The phonon spectral density predicts
 two excitation peaks, one due to SC gap and the other due to JT distortion.
 The DJT coupling suppresses the JT gap while it enhances SC gap as well as
 SC transition temperature. The evolution of these excitation peaks are 
investigated by varying different model parameters of the system.\end{abstract}


\section{Introduction}
The structural transition plays dominant role on various physical 
properties as well as on the occurrence of 
superconductivity in some of the 
high-$T_{c}$ cuprates.  The  system $La_{1-x}Ba_{x}CuO_{4}$ (LBCO)
with $x=0.15$ shows a phase transition from high temperature 
tetragonal to low temperature orthorhombic phase  
at $T_{d}=180K$ accompanied by the SC phase 
transition at $35K$ at the onset of superconductivity.
The neutron diffraction on LBCO shows an orthorhombic phase below 
$T=180K$\cite{rf1,rf2,rf3}. The orthorhombic structure of the system 
$La_{2-x}Sr_{x}CuO_{4}$ (LSCO) changes near the SC transition temperature 
$T_{c}\simeq36K$, but a structural transition takes place at 
$T_{d}=220K$\cite{rf4}. The thermal expansion 
measurements on the cuprate systems of Lang $et~al$\cite{rf3} 
demonstrates that the high-T$_c$ is observed at optimum doping for
 which the system undergoes the tetragonal to orthorhombic transition. The 
structural transition associated with the lowering of the crystallographic 
symmetry has also been observed in other high-$T_{c}$ cuprates 
\cite{rf5,rf6,rf7}.  Besides the system LSCO the other high-T$_c$
 superconductors (HTSCs) also exhibit similar anomalies in the
 isotope effect exponent. A correlation between the superconducting 
transition temperature and the scattering intensity from 
$43cm^{-1}$ and $100cm^{-1}$ modes for bismuthates are 
studied by Sugai $et~al$\cite{rf7a}.

The experimental data shows that the isotope shift exponent ($\alpha$)
 of HTSCs shows the influence of structural phase transition. It is 
observed that the transition temperature T$_c$ is enhanced with the 
increase of dopant concentration, while $\alpha$ decreases monotonically.
 The experiments on LSCO by Crowford $et~al$\cite{rf8,rf9} gives the
 higher value of $\alpha$ than the BCS value of $\alpha =0.5$ calculated
 from the phonon mechanism of pairing. It is generally noticed that value
 of isotope exponent $\alpha$ is smaller than 0.5. The LSCO system gives 
a highest SC transition temperature of T$_c^{max}=38K$ at optimum doping. 
Further, it is observed that the T$_c$ decreases with the decrease of 
dopant concentration while the value of $\alpha$ increases from 0.1 to 0.6.
 These variations can be understood if the system undergoes a structural
 phase transition.

The band Jahn-Teller (BJT) distortion 
usually induces structural phase transition.
 The lattice strain removes the degeneracy in the conduction electron 
states by lowering the lattice symmetry. The system is stabilized due 
to the gain in electronic energy at the cost in the elastic energy. In
 the strained state, the conduction band is maximally occupied. For large
 distortion, the gap due to BJT distortion produces an insulating phase. 
In this scenario the doping of the holes modifies the occupation 
probabilities of the two bands.

For the first time Raman scattering exhibit SC gap excitation mode in 
the layered system 2H-NbSe$_2$\cite{rf10,rf11}, where Raman peak has 
coupled to the charge density wave (CDW) amplitude mode of the system.
 They have observed SC gap excitation mode at $16cm^{-1}$ and CDW amplitude
 mode at $40cm^{-1}$ in the Raman spectra due to the coupling of SC gap to
 the CDW amplitude phonon. Several authors\cite{rf12,rf13,rf14,rf15} have
 tried to explain the origin of the collective modes of the SC and CDW 
state appearing in Raman spectra. Recently Rout $et~al$\cite{mod2} have 
reported the interplay of JT distortion and superconductivity into Raman
 active excitation peaks. It is to note that the authors have considered 
only static JT distortion in their calculation. More recently Raj 
$et~al$\cite{Raj} have considered the interplay of superconductivity
 and static JT distortion in presence of a DJT distortion and reported 
the results for modified BCS type gap equation which is strongly 
influenced by DJT distortion. Based on the same model\cite{Raj} we 
calculate here phonon Green's function and report Raman active excitation
 peaks which can explain the Raman scattering observed in high-T$_c$ cuprates,
 layered 2H-NbSe$_2$\cite{rf10,rf11} as well as the high-T$_c$ system
 like bismuthates [$K_xBa_{1-x}BiO_3$]\cite{rf7a}. The rest of the work is
 organized as the theoretical model in section 2, the phonon self-energy 
is calculated in section 3, the electron response function in section 4, 
Raman spectral intensity is presented in section 5, the results are discussed 
in section 6 and finally concluded in section 7. 

\section{Theoretical Model}
The present model study attempts to investigate the effect of 
dynamic Jahn-Teller distortion on the superconducting  gap in high-T$_c$ 
superconductors. In order to introduce JT effect in the conduction band
 we consider two orbitals ($\alpha = 1,~2$) which was splitted due to the
 introduction of the JT effect. In order to represent SC, static JT and 
DJT effect we write the Hamiltonian below based upon our earlier 
models\cite{Raj,mod2,mod1,mod3}.
\begin{equation}
H_c=\sum_{k\sigma}\epsilon_{k}\left(c^{\dagger}_{1k\sigma}
 c_{1k\sigma}+c^{\dagger}_{2k\sigma}c_{2k\sigma}\right)
\label{ee1}
\end{equation}
\begin{equation}
H_{e-L}=Ge~\sum_{k\sigma}\left(c^\dagger_{1k\sigma}c_{1k\sigma}
- c^\dagger_{2k\sigma}c_{2k\sigma}\right)
\label{ee2}
\end{equation}
The Hamiltonian $H_c$ represents the hopping of the electrons between 
the two nearest neighbors for the two degenerate orbitals designated 
as 1 and 2. The dispersion of the degenerate band in a two dimensional
 CuO$_2$ plane is written as $\epsilon_k=-2t_0(\cos k_x+\cos k_y)$.
 Here $c^{\dagger}_{\alpha k\sigma}(c_{\alpha k\sigma})$, for $\alpha=1$ 
and 2, are creation (annihilation) operators of the conduction electrons
 of copper ions for two orbitals with momentum $k$ and spin $\sigma$. 
The Hamiltonian $H_{e-L}$ represents the static JT interaction where $G$
 is the strength of the electron-lattice interaction and $e$ is the strength 
of the isotropic static lattice strain. The lattice strain splits the 
single degenerate band into two bands with energies 
$\epsilon_{1k,2k}=\epsilon_{k}\pm{Ge}$. The elastic energy of the system is
 $\frac{1}{2}Ce^{2}$ with $C$ representing the elastic constant. The 
minimization of the free energy of the electron including the
 elastic energy helps to find the expression for 
lattice strain. It is shown earlier \cite{mod2,mod1,mod3} that the static 
lattice strain suppresses the SC gap parameter in the interplay region of 
the two order parameters.

In order to investigate the phonon response in the HTSCs,
  we consider the phonon interaction to the density
 of the conduction electrons of both the orbitals as well as the phonon 
coupling to the difference in electron densities of the JT distorted 
 orbitals of the conduction band. The electron-phonon interaction 
Hamiltonian is written as 
\begin{eqnarray}
H_{e-p}&=&\sum_{\alpha,k,\sigma}f_1(q)\left(c^{\dagger}_{\alpha, k+q, \sigma}
c_{\alpha, k, \sigma}\right)A_{q}-\sum_{\alpha,k,\sigma}(-1)^{\alpha}f_2(q)e\left(c^{\dagger}_{\alpha, k+q,\sigma}
c_{\alpha, k, \sigma}\right)A_{q}\nonumber \\
&=&\sum_{\alpha, k,\sigma}s_{\alpha}(q)\left(c^{\dagger}_{\alpha, k+q, \sigma}
c_{\alpha, k, \sigma}\right)A_{q}.
\label{ee3}
\end{eqnarray}
The strength of the electron-phonon coupling $s_{\alpha}(q)$ is defined as
 $s_{\alpha}=f_1(q)-(-1)^{\alpha}f_2(q)e$ in which $f_1(q)$ is 
the normal EP coupling and $f_2(q)$ is the dynamic Jahn-Teller EP coupling. 
The $q^{th}$-Fourier component of the phonon displacement operator is
 $A_{q}=b_{q}+b_{-q}^{\dagger}$ with $b_{q}^{\dagger}$ ($b_{q}$) 
defining the phonon  creation (annihilation) operator for wave vector $q$. 
Further the free phonon Hamiltonian $H_p$ is given in harmonic approximation as
\begin{equation}
H_p=\sum_{q}\omega_{q}b^{\dagger}_{q}b_{q},
\label{ee4}
\end{equation}
with $\omega_{q}$ being the free phonon frequency.

Our main objective in the present report is to study the effect of
 dynamic JT distortion on the superconductivity in HTSCs. 
The $d$-wave models have gained substantial support recently
over $s$-wave pairing as the mechanism by which high temperature
superconductivity might be explained. The establishment of $d$-wave
symmetry in cuprates does not necessarily specify a high-T$_c$
 mechanism. It does not impose well defined constraints on possible
models for this mechanism. While the spin fluctuation pairing
mechanism leads naturally to an ordered parameter with $d$-wave
symmetry, the conventional BCS electron phonon pairing interaction
give rise to $s$-wave superconductivity. M$\ddot{u}$ller\cite{Muller1,Muller2}
 has considered $d$-wave symmetry near the surface and $s$-wave symmetry 
in bulk of cuprate superconductors and applied them to small angle neutron
 scattering results. Zhao\cite{Zhao} has concluded from angle-resolved
 photo-emission spectroscopy and scanning tunneling spectroscopy that
 the superconductivity is extended $s$-wave rather than $d$-wave. So
 it is not definite that the bulk superconductivity  is $d$-wave.
We consider here
 the $s$-wave pairing interaction within the same orbitals and the same 
strength of interactions for the two orbitals. The BCS type pairing 
Hamiltonian is considered here for the two orbitals as
\begin{equation}
H_I=-\Delta\sum_{\alpha, k}\left(c_{\alpha, k,\uparrow}^{\dagger}c_{\alpha, -k,\downarrow}^{\dagger}+c_{\alpha, -k,\downarrow}c_{\alpha, k,\uparrow}\right).
\label{ee5}
\end{equation}
In order to simulate an attractive interaction to produce Cooper pairs,
 the energy dependence of the interaction potential is taken as
\begin{equation}
U(\epsilon)=U_0\left[1-\frac{(\epsilon-\epsilon_F)^4}{\omega_{D}^{4}} \right]^{\frac{1}{2}},
\label{ee6}
\end{equation}
where $U_0$ is the effective attractive Coulomb interaction, $\omega_D$ 
is cut-off energy and $\epsilon_F$ represents the Fermi energy. 
In order to produce band splitting due to JT distortion, we consider an 
energy dependent density of state $N(\epsilon)$ around the center of the
 conduction band in the system. Such a logarithmic model density
 of state\cite{mod3} is given by
\begin{equation}
N(\epsilon)=N(0)\sqrt{1-|{\frac{\epsilon}{D}}|}ln|{\frac{D^2}{\epsilon^2}}|,
\label{ee7}
\end{equation}
where $2D=W$ is the conduction band width. The total Hamiltonian 
describing the DJT effect and SC interaction in high-T$_c$ 
cuprate systems can be written as
\begin{equation}
H=H_c+H_{e-L}+H_I+H_{e-p}+H_p.
\label{ee8}
\end{equation}
The phonon induced superconductivity given in eqn.(\ref{ee5}) is the 
consequence of the electron-phonon interaction. Here we have included 
the phonon mediated superconducting pairing as well as the normal 
electron-phonon interaction and both the static and dynamic band 
Jahn-Teller interactions in the total Hamiltonian in order to study
 their effect on the superconducting gap. The appearance of the Debye
 frequency $\omega_D$ in the energy dependence of repulsive Coulomb 
energy in eqn.(\ref{ee6}) implies that the pairing interaction is
 mediated via phonons. The electron-phonon interaction which is 
considered explicitly in eqn.(\ref{ee3}) is just a very small interaction 
contributing negligibly to the pairing. However, the main interest in the
 present work is to investigate the effect of Jahn-Teller strain on the 
pairing and the superconducting gap as well.
     
\section{Phonon self-energy for finite $q$}
The phonon self-energy for the system is evaluated by the double-time 
Green's function technique of Zubarev\cite{Zubarev} using the equation
 of motion method.
 The single particle phonon Green's function is defined as
\begin{eqnarray}
D_{qq^{\prime}}(t~-~t^{\prime})~=~\big<\big<~A_{q}(t);A_{q}(t^{\prime})\big>\big>~\nonumber \\
\equiv-i\theta(t~-~t^{\prime})~\big<[~A_{q}(t);A_{q^{\prime}}(t^{\prime})]\big>.
\label{eq13}
\end{eqnarray}
Using the total Hamiltonian of eqn.(\ref{ee8}), the Fourier transformed phonon Green's function reduces to
\begin{equation}
 D_{qq^{'}}(\omega)=\delta_{-qq^{'}}D^{0}_{q}(\omega)+2\pi s^{2}_\alpha D^{0}_{q}(\omega)\chi_{qq^{'}}(\omega)D^{0}_{q^{'}}(\omega),
\label{ee10}
\end{equation}
where the Fourier transformed free phonon propagator is given by
\begin{equation}
D^{0}_{q}(\omega)=\frac{\omega_{q}}{\pi(\omega^{2}-\omega_{q}^{2})}.
\label{ee 5}
\end{equation}
Applying Dyson's approximation, the eqn.(\ref{ee10}) can 
be written in a closed form as
\begin{equation}
D_{qq^{'}}(\omega)=\frac{1}{\pi}\frac{\omega_{q}}{[\omega^{2}-\omega_{q}^{2}-\sum(\omega, q)]},
\label{ee 6}
\end{equation}
where the phonon self-energy appears as
\begin{equation}
\sum(\omega, q)=4\pi \omega_q \chi_{qq^\prime}(\omega),
\label{ee 7}
\end{equation}
 $\chi_{qq^{\prime}}(\omega)$ is the electron density response function, 
which consists of the contributions from the two degenerate orbitals 1 and 2
as $\chi_{qq^{'}}(\omega)=\chi^{1}_{qq^{'}}(\omega)+\chi^{2}_{qq^{'}}(\omega)$
and the electron-phonon coupling parameters $s_\alpha=f_1\pm f_2\times {e}$.
The two response functions corresponding to the two orbitals are given by
\begin{equation}
\chi^{1}_{qq^{'}}(\omega)=\sum_{k\sigma k^{'}\sigma^{'}}s^2_{\alpha}\Gamma_{1}(k,\sigma,k^{'},\sigma^{'},q,q^{'},\omega),
\end{equation}
\begin{equation}
\chi^{2}_{qq^{'}}(\omega)=\sum_{k\sigma k^{'}\sigma^{'}}s^2_{\alpha}\Gamma_{2}(k,\sigma,k^{'},\sigma^{'},q,q^{'},\omega),
\end{equation}
with
\begin{equation}
\Gamma_{1}(k,q,\sigma,\omega)=\Gamma^{a}_{1}(k,-q,\omega)+\Gamma^{b}_{1}(k,-q,\omega),
\end{equation}
\begin{equation}
\Gamma_{2}(k,q,\sigma,\omega)=\Gamma^{a}_{2}(k,-q,\omega)+\Gamma^{b}_{2}(k,-q,\omega).
\end{equation}
Further the two particle Green's functions $\Gamma_1$ and $\Gamma_2$ contain 
the other Green's functions defined as
\begin{equation}
\Gamma^{a}_{1}(k,-q,\omega)=\Big<\Big<\alpha^{a}_{k};\chi_{1k^{'}}\Big>\Big>_{\omega},\Gamma^{b}_{1}(k,-q,\omega)=\Big<\Big<\alpha^{b}_{k};\chi_{1k^{'}}\Big>\Big>_{\omega},
\end{equation}
\begin{equation}
\Gamma^{a}_{2}(k,-q,\omega)=\Big<\Big<\alpha^{a}_{k};\chi_{2k^{'}}\Big>\Big>_{\omega},\Gamma^{b}_{2}(k,-q,\omega)=\Big<\Big<\alpha^{b}_{k};\chi_{2k^{'}}\Big>\Big>_{\omega}.
\end{equation}
The new two particle Green's functions are expressed in terms of the electron 
density operators like $\alpha_{k}^{a},~\alpha_{k}^{b},~\beta_{k}^{b},~\beta_{k}^{b},~X_{1k^{\prime}},~X_{2k^{\prime}}$, which are defined below:
\begin{eqnarray}
&&\alpha^{a}_{k}=C^{\dagger}_{1k-q\uparrow}C_{1k\uparrow},\alpha^{b}_{k}=C^{\dagger}_{1k-q\downarrow}C_{1k\downarrow},\nonumber\\
&&\beta^{a}_{k}=C^{\dagger}_{2k-q\uparrow}C_{2k\uparrow},\beta^{b}_{k}=C^{\dagger}_{2k-q\downarrow}C_{2k\downarrow},\\ \nonumber
&&\chi_{1k^{'}}=C^{\dagger}_{1k^{'}-q^{'},\sigma^{'}}C_{1k^{'}\sigma^{'}},
\chi_{2k^{'}}=C^{\dagger}_{2k^{'}-q^{'},\sigma^{'}}C_{2k^{'}\sigma^{'}}.
\end{eqnarray}
\section{Electron response function for finite $q$}
For the calculation of the two particle 
Green's function $\Gamma_1(k-q,\omega)$ for
 the JT split orbital 1, the other higher order Green's functions involving
 superconducting and normal state operators are defined as
\[A^{a}_{1}(k,\omega)=\left<\left<\gamma^{1}_{k};X_{1k^{\prime}}\right>\right>_{\omega},~~ A^{b}_{1}(k,\omega)=\left<\left<\gamma^{2}_{k};X_{1k^{\prime}}\right>\right>_{\omega},\]
\[A^{a}_{2}(-k+q,\omega)=\left<\left<\gamma^{1}_{-k+q};X_{1k^{\prime}}\right>\right>_{\omega},~~A^{b}_{2}(-k+q,\omega)=\left<\left<\gamma^{2}_{-k+q};X_{1k^{\prime}}\right>\right>_{\omega},\]
\[A^{a}_{3}(-k+q,\omega)=\left<\left<\alpha^{a}_{-k-q};X_{1k^{\prime}}\right>\right>_{\omega},~~A^{b}_{3}(-k+q,\omega)=\left<\left<\alpha^{b}_{-k+q};X_{1k^{\prime}}\right>\right>_{\omega}.\]
The SC Green's functions $A_1,~A_2,~A_3$ involve the two particle Cooper
 pairing operators as defined below
\begin{eqnarray}
&&\gamma^{1}_{k}=C^{\dagger}_{1,k-q\uparrow}C^{\dagger}_{1,-k\downarrow},~~~~\gamma^{2}_{k}=C_{1,-k+q\downarrow}C_{1,k\uparrow},\nonumber \\
&&\gamma^{1}_{-k+q}=C^{\dagger}_{1,-k\uparrow}C^{\dagger}_{1,k-q\downarrow},~~~~\gamma^{2}_{-k+q}=C_{1,k\downarrow}C_{1,-k+q\uparrow}.
\label{ee27}
\end{eqnarray}
The new Green's functions are related to each other as
\begin{eqnarray}
&&A_{1}(k,\omega)=A^{a}_{1}(k,\omega)-A^{b}_{1}(k,\omega),\nonumber\\
&&A_{2}(-k+q,\omega)=A^{a}_{2}(-k+q,\omega)-A^{b}_{2}(-k+q,\omega),\nonumber\\
&&A_{3}(-k+q,\omega)=A^{a}_{3}(-k+q,\omega)-A^{b}_{3}(-k+q,\omega).
\label{ee28}
\end{eqnarray}
Similarly for the calculation of the two particle Green's function
 $\Gamma_2(k-q,\omega)$ corresponding to the orbital 2, the other 
higher order Green's functions involving superconducting and normal
 state operators are defined as
\begin{equation}
B^{a}_{1}(k,\omega)=\left<\left<\delta^{1}_{k};X_{2k^{\prime}}\right>\right>_{\omega},~~B^{b}_{1}(k,\omega)=\left<\left<\delta^{2}_{k};X_{2k^{\prime}}\right>\right>_{\omega},
\label{ee29}
\end{equation}
\begin{equation}
B^{a}_{2}(-k+q,\omega)=\left<\left<\delta^{1}_{-k+q};X_{2k^{\prime}}\right>\right>_{\omega},~~B^{b}_{2}(-k+q,\omega)=\left<\left<\delta^{2}_{-k+q};X_{2k^{\prime}}\right>\right>_{\omega},
\label{ee30}
\end{equation}
\begin{equation}
B^{a}_{3}(-k+q,\omega)=\left<\left<\alpha^{a}_{-k+q};X_{2k^{\prime}}\right>\right>_{\omega},~~B^{b}_{3}(-k+q,\omega)=\left<\left<\alpha^{b}_{-k+q};X_{2k^{\prime}}\right>\right>_{\omega}.
\label{ee31}
\end{equation}
The SC Green's functions $B_1,~B_2,~B_3$ involve the two particle Cooper pairing operators as defined below
\begin{eqnarray}
&&\delta^{1}_{k}=C^{\dagger}_{2,k-q\uparrow}C^{\dagger}_{2,-k\downarrow},~~~~\delta^{2}_{k}=C_{2,-k+q\downarrow}C_{2,k\uparrow},\nonumber\\
&&\delta^{1}_{-k+q}=C^{\dagger}_{2,-k\uparrow}C^{\dagger}_{2,k-q\downarrow},~~~~\delta^{2}_{-k+q}=C_{2,k\downarrow}C_{2,-k+q\uparrow}.
\end{eqnarray}
The new Green's functions defined in equations(\ref{ee29})-(\ref{ee31}), are again related as
\begin{eqnarray}
&&B_{1}(k,\omega)=B^{a}_{1}(k,\omega)-B^{b}_{1}(k,\omega),\nonumber\\
&&B_{2}(-k+q,\omega)=B^{a}_{2}(-k+q,\omega)-A^{b}_{2}(-k+q,\omega),\nonumber\\
&&B_{3}(-k+q,\omega)=B^{a}_{3}(-k+q,\omega)-A^{b}_{3}(-k+q,\omega).
\label{ee31a}
\end{eqnarray}

~~~The electron response functions $\chi_{qq^{\prime}}^{1}(\omega)$ and 
$\chi_{qq^{\prime}}^{2}(\omega)$ are calculated for the two orbitals 
separately and the two are added together to get the final result. The
 frequency $(\omega)$ and wave vector ($q$) dependent total electron response 
function at finite temperature then can be written as
\begin{eqnarray}
\chi_{qq}(\omega)&=&\chi^{1}_{qq}(\omega)+\chi^{2}_{qq}(\omega)\nonumber\\
&=&\frac{1}{2\pi}\sum_{k\alpha}\frac{1}{|D_{\alpha}|}[(\omega^{2}-\epsilon^{2}_{\alpha +}(k,q))(\omega-\epsilon_{\alpha -}(k,q))\nonumber\\
&\times&(n_{\alpha,k-q}-n_{\alpha,k})+4\omega\Delta(\omega-\epsilon_{\alpha-}(k,q))(\phi_{\alpha,k-q}+\phi_{\alpha,k})],
\end{eqnarray}
 where $\alpha=1$ and $2$, $n_{\alpha,k-q}=n_{\alpha,k-q\uparrow}+n_{\alpha,k-q\downarrow}$ and the denominator
 \begin{equation}
 |D_{\alpha}|=(\omega^{2}-\tilde{E}^{2}_{\alpha+}(k,q))(\omega^{2}-\tilde{E}^{2}_{\alpha-}(k,q))
 \end{equation}
 with $\tilde{E}_{\alpha,\pm}(k,q)=\tilde{E}_{\alpha, k-q}\pm \tilde{E}_{\alpha, k}, ~~\epsilon_{\alpha\pm}(k,q)=\epsilon_{\alpha,k-q)}\pm \epsilon_{\alpha, k}$ and $\tilde{E}^{2}_{\alpha,k}=(\tilde{\epsilon}^{2}_{\alpha,k}+\tilde{\Delta}_{\alpha}^{2})$ 
are the energy of the SC gap excitation, with 
 $\epsilon_{1k-q}=(\epsilon_{k-q}+Ge)$ and 
 $\epsilon_{2k-q}=(\epsilon_{k-q}-Ge)$. In presence of normal EP coupling 
and the DJT coupling, the renormalized conduction band dispersion 
$\tilde{\epsilon}_{\alpha, k}$ and the SC gap $\tilde{\Delta}_{\alpha}$ are
 given in equations(\ref{ee15}) and (\ref{ee16}) respectively. The SC 
amplitudes $\phi_{\alpha,k}$ are given by
\begin{equation}
 \phi_{\alpha,k}=\left<C^{\dagger}_{\alpha,k\uparrow} C^{\dagger}_{\alpha,-k\downarrow}\right>=\frac{\tilde{\Delta}_{\alpha}}{2\tilde{E}_{\alpha,k}}\tanh\Big(\frac{\beta \tilde{E}_{\alpha,k}}{2}\Big).
\label{ee35}
 \end{equation}
The SC gap ($\Delta$) and the lattice strain due to JT distortion ($e$) are 
temperature dependent quantities which can be determined self-consistently.

In order to calculate the Raman spectrum, it is necessary to evaluate the 
electron response function in the limit $q\rightarrow 0$, at a finite 
temperature ($T\neq~0$). The electron response function in the limit 
$q=0$ reduces to
 \begin{equation}
\chi(\omega,q=0)=\frac{1}{2\pi}\sum_{\alpha,k}\left[\frac{8\tilde{\Delta}_{\alpha}\phi_{\alpha,k}}{\omega^2-4\tilde{E}^2_{\alpha,k}}\right],
\label{ee36}
 \end{equation}
where $\tilde{E}^2_{\alpha,k}=\tilde{\epsilon}_{\alpha,k}^2+\tilde{\Delta}_{\alpha}^2$ with
\begin{equation}
\tilde{\epsilon}_{\alpha k}=\epsilon_{\alpha k}+s^{2}_{\alpha}\left[\frac{(\omega-\epsilon_{\alpha k})N_0}{(\omega-\epsilon_{\alpha k})^{2}+\Delta^{2}}\right],
\label{ee15}
\end{equation}
and
\begin{equation}
\tilde{\Delta}_{\alpha k}=\Delta+{s^{2}_{\alpha}}\left[\frac{\Delta N_0}{(\omega-\epsilon_{\alpha k})^{2}+\Delta^{2}}\right]
\label{ee16}
\end{equation}
where
\[N_0=2(e^{\beta\omega_0}-1)^{-1},\] with $\beta=1/k_BT$ 
and $\omega_0$ being the bare phonon frequency at temperature T.
\begin{equation}
\chi(\omega+i\eta,q=0)=\frac{1}{2\pi}\sum_{\alpha,k}8\tilde{\Delta}_{\alpha}\phi_{\alpha,k}\left[\frac{\omega^2-4\tilde{E}^2_{\alpha,k}}{|D_{\alpha 0}|}-i\frac{2\eta \omega}{|D_{\alpha 0}|}\right]
\label{ee37}
\end{equation}
\begin{equation}
\sum(\omega+i\eta,q=0)=16\omega_0\sum_{\alpha,k}\tilde{\Delta}_{\alpha}s_{\alpha}^2\phi_{\alpha,k}\left[\frac{\omega^2-4\tilde{E}^2_{\alpha,k}}{|D_{\alpha 0}|}-i\frac{2\eta \omega}{|D_{\alpha 0}|}\right]
\label{ee38}
\end{equation}
where $|D_{\alpha 0}|=(\omega^2-4\tilde{E}_{\alpha,k}^2)^2+4\eta^2\omega^2$.

The temperature dependent SC gap ($\Delta$) and lattice strain ($e$) 
are calculated by using the total Hamiltonian given in equations(\ref{ee1})
 to (\ref{ee4}). The calculations within a mean-field approximation give 
rise to a modified BCS type gap equation with involving renormalized 
conduction band dispersion $\tilde{\epsilon}_{\alpha k}$ and the renormalized
 SC gap parameter $\tilde{\Delta}_{\alpha}$ has written in 
equations(\ref{ee15}) and (\ref{ee16}). The SC gap equation and the 
lattice strain are calculated in terms of these renormalized
 quantities and the equations are
\begin{equation}
1=\int_{-\omega_{D}}^{\omega_{D}}U(\epsilon)N(\epsilon)d\epsilon_{k}\left[\frac{1}{2\tilde{E}_{1k}}
\tanh\left(\frac{1}{2}\beta\tilde{E}_{1k}\right)+\frac{1}{2\tilde{E}_{2k}}\tanh\left(\frac{1}{2}\beta\tilde{E}_{2k}\right)\right],
\label{ee39}
\end{equation}
\begin{equation}
e=\left(\frac{-G}{C_{0}}\right)\int_{-W/2}^{W/2}N(\epsilon)d\epsilon_{k}\left[\frac{\tilde{\epsilon}_{1k}}{2\tilde{E}_{1k}}\tanh\left(\frac{1}{2}\beta\tilde{E}_{1k}\right)-\frac{\tilde{\epsilon}_{2k}}{2\tilde{E}_{2k}}\tanh\left(\frac{1}{2}\beta\tilde{E}_{2k}\right)\right].
\label{ee40}
\end{equation}
These two gap equations are solved numerically and self-consistently
 and reported earlier\cite{Raj}. The values of SC gap $\Delta$ and
 lattice strain $e$ are calculated on the numerical solution at a 
particular temperature and later on used for the calculation of Raman
 spectra at a given finite temperature.

\section{Raman spectra in the limit $q=0$ at finite $T$}
The Raman intensity is given by the spectral density function (SDF) 
of the zone center phonon. The SDF in general is defined by 
\begin{equation}
S(\omega,q)=-\pi Im D_{qq^{\prime}}(\omega)|_{\eta \rightarrow 0}
\label{ee42}
\end{equation}
where the phonon Green's function $D_{qq^{\prime}}(\omega)$ is given in
 equation(\ref{ee 6}). At the limit $q=0$ and finite temperature (T) we have 
\begin{equation}
D_{00}(\omega +i\eta)=\frac{1}{\pi}\left[\frac{\omega_0(A_1-iB_1)}{A_1^2+B_1^2}\right],
\label{ee42b}
\end{equation}
and hence
\begin{equation}
S(\omega,q=0)=\frac{\omega_{0}B_{1}}{A^{2}_{1}+B^{2}_{1}}
\label{ee43}
\end{equation}
 where $A_{1}$ and $B_{1}$ are
 \begin{equation}
 A_{1}=\omega^{2}-\omega^{2}_{0}-A_2,~~~ B_{1}=2 \eta\omega-B_2,
 \end{equation}
 with
 $A_2=8\omega_{0}\sum_{\alpha}s_{\alpha}^2\int N(0)r d\epsilon_{k}G_{\alpha}$ and $B_2=-16\omega_{0}\eta \sum_{\alpha}s_{\alpha}^2\int N(0)r d\epsilon_{k}H_{\alpha}$
where
  \begin{equation}
 G_{\alpha}=\left[\frac{\tilde{\Delta}_{\alpha}^{2}(\omega^{2}-4\tilde{E}^{2}_{\alpha k})}{\tilde{E}_{\alpha k}|D_{\alpha 0}|}\tanh\left(\frac{\beta \tilde{E}_{\alpha,k}}{2}\right)\right],
   \end{equation}
   \begin{equation}
H_{\alpha}=\left[\frac{\tilde{\Delta}_{\alpha}^{2}}{\tilde{E}_{\alpha k}|D_{\alpha 0}|(\omega)}\tanh\left(\frac{\beta \tilde{E}_{\alpha,k}}{2}\right)\right].
   \end{equation}
 The different physical quantities of the atomic subsystem are made
 dimensionless dividing them by the hopping integral $2t_{0}$, the width
 of the conduction band is $W=8t_{0}$. The dimensionless parameters are
 the SC gap parameter $z=\frac{\Delta}{2t_{0}}$, the SC coupling parameter
 $g=N(0)U_{0}$, the Debye frequency ${\omega_{D}}=\frac{\omega_{D}}{2t_{0}}$,
 the reduced temperature $t=\frac{k_{B}T}{2t_{0}}$, the reduced 
 incident photon frequency $c_1=\frac{\omega}{2t_0}$,
 the phonon vibrational frequency 
$\omega_1=\frac{\omega_q}{2t_0}$, the reduced bare phonon frequency 
$p=\frac{\omega_0}{2t_0}$, $x=\frac{\epsilon_{k}}{2t_{0}}$, 
$e_q=\frac{qv_f}{2t_0}$, the JT coupling constant 
$g_1=\frac{G}{2t_{0}}$, the reduced lattice strain $\tilde{e}=\frac{e}{2t_0}$,
 the normal electron-phonon coupling $\lambda_1=\frac{f_1}{2t_{0}}$, the
 dynamic electron-phonon coupling $\lambda_2=\frac{f_2}{2t_{0}}$, 
  spectral width $e_1=\frac{\eta}{2t_0}$ and 
we have taken the JT distortion energy ${e}^\prime=g_1\times \tilde{e}$.

\section{Results and Discussion}
   Before calculating the spectral density function (SDF) under dynamic
 condition of JT effect at a finite temperature, we solved the SC gap 
parameter ($z$) and lattice strain ($e^\prime$). The self-consistent solution 
of these two temperature dependent parameters are plotted in figure 1. 
Earlier the solution for $z$ and $e^\prime$ are solved self-consistently under
 static and dynamic limits of JT effect and the results are 
reported\cite{Raj}. It has been observed that the lattice strain is 
suppressed throughout the temperature range under the DJT condition and
 the SC gap ($z$) is enhanced accompanied by an 
enhancement in transition temperature $t_c$. However, the temperature 
dependencies of these two parameters in static limit are similar to their
 counter part in the dynamic limit. It is concluded from this that at a 
given temperature the SC gap in dynamic limit is larger than its value 
in static limit, whereas the magnitude of the lattice strain has lesser 
value in dynamic limit than in static limit. Accordingly the Raman 
excitation peaks in dynamic limit is expected to appear at slightly 
displaced position than that of the static limit. The evolution of 
these Raman peaks will be studied in the present investigation by varying
 the model electronic as well as lattice parameters of the system as shown
 in subsequent figures from 2 to 9. In the present case figure 1 shows 
the temperature dependent SC gap ($z$) and lattice strain ($e^\prime$) under
 dynamic limit of the DJT effect. In this case the SC transition temperature
 appears at $t_c\simeq 0.0052$ and the lattice distortion temperature
 appears at $t_d\simeq 0.0082$. It is to note further that lattice strain
 is suppressed at lower temperatures where the superconductivity coexists
 with lattice distortion.

\vspace{0.2in}

%%%%%%%%%%%%%%%%%%%%%%%%%%%%%%%%%%%%%%%%%%%%%%%%%%%%%%%%%%%%%%%%%%%%%%%%%%%%%
\bigskip

{\vbox{
\begin{center}
      \epsfig{file=gap.eps,width=8cm,height=6cm}
      \vspace{-1.0cm}
 %     \begin{figure}[p]
  %    \label{c1f1}
   %   \vspace{0.5cm}
    %  \end{figure}
\end{center}
}}

\vspace{0.5cm}

{\small {\bf Fig.} \ {\bf 1} \
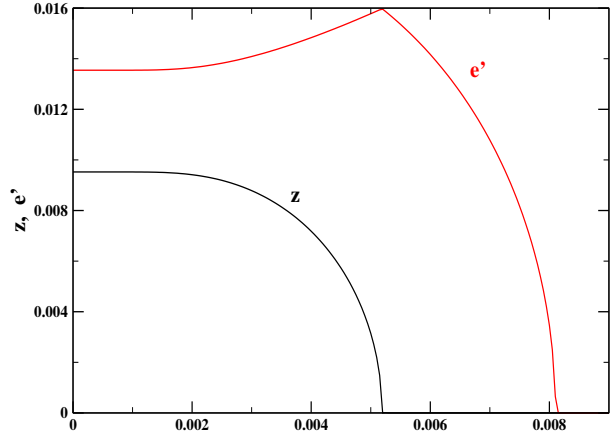
The self-consistent plots of SC gap $z$ and JT gap energy
 ${e}^{\prime}$ vs. reduced temperature $t$ for
 the SC coupling $g=0.031$, the static JT coupling $g_1=0.152$, the phonon
 vibrational frequency $\omega_1=0.06$, the incident photon
 frequency $c_1=0.1$, the normal EP
 coupling $\lambda_1=0.14$ and the DJT coupling $\lambda_2=0.10$.}
%%%%%%%%%%%%%%%%%%%%%%%%%%%%%%%%%%%%%%%%%%%%%%%%%%%%%%%%%%%%%%%%%%%%%%%%%%%%%%%

The Raman spectral intensity or the phonon SDF is
 plotted in figure 2 for a given temperature $t=0.001$, where the 
superconductivity and lattice strain coexist in the system. In absence
 of normal EP coupling i.e., $\lambda_1 =0$, the phonon self-energy 
($\sum(\omega, q)=0$) is zero. Hence there appears a peak $p_0$ centered 
at $\tilde{\omega}=\frac{\omega}{\omega_0}=1$ corresponding to the bare
 phonon frequency $(p=\frac{\omega_0}{2t_0}=1)$ for the long wave length
 optical phonon ($q=0$). For a finite normal EP coupling $(\lambda_1=0.14)$
 and DJT coupling $(\lambda_2=0.10)$ at finite temperature $t=0.001$ with given
 SC gap $(z=0.00952)$ and lattice strain $(e^\prime=g_1\times e \simeq 0.0136)$,
 the position of the bare phonon peak $p_0$ shifts to the higher
 frequencies i.e., $\tilde{\omega}\simeq 1.05$ indicating the hardening
 behavior of the phonon frequency due to renormalization effect. For these
 two given values of $\lambda_1$ and $\lambda_2$ there appears two new Raman 
active peaks $p_1$ and $p_2$ at lower frequencies. The low frequency 
peak $p_1$ appears at energy $\tilde{\omega}\simeq 0.125$. This corresponds
 to the excitation energy of the peak i.e., 
$c_1=p\tilde{\omega}\simeq 0.08\times 0.125=0.010$. This energy is slightly
 higher than the SC gap energy $\Delta \simeq 0.00952$ at $t=0.001$ as
 determined from figure 1. The slightly higher value obtained in the 
spectral function in figure 2 is due to other interaction including the
 DJT effect. Thus the peak $p_1$ arises due to the SC excitation gap. 
The second peak $p_2$ appearing at the frequency $\tilde{\omega}\simeq 0.245$, 
 corresponds to an excitation energy of the peak i.e., 
$c_1=p\tilde{\omega}\simeq 0.08\times 0.245=0.0196$. This energy is 
slightly higher than, but close to the gap associated with JT distortion 
at $t=0.001$ which is given by $e^{\prime}=g_1\times e=0.0136$. So the 
peak $p_2$ appearing in the SDF can be attributed to the phonon excitation 
arising due to lattice distortion as the system undergoes a structural 
phase transition. The interplay of these two order parameters SC gap $(z)$ 
and JT energy ($e^{\prime}$) is investigated below from figure 3 to 9.

\vspace{0.2in}

%%%%%%%%%%%%%%%%%%%%%%%%%%%%%%%%%%%%%%%%%%%%%%%%%%%%%%%%%%%%%%%%%%%%%%%%%%%%%
\bigskip

{\vbox{
\begin{center}
      \epsfig{file=rmn0.eps,width=8cm,height=6cm}
      \vspace{-1.0cm}
 %     \begin{figure}[p]
  %    \label{c1f1}
   %   \vspace{0.5cm}
    %  \end{figure}
\end{center}
}}

\vspace{0.5cm}

{\small {\bf Fig.} \ {\bf 2} \
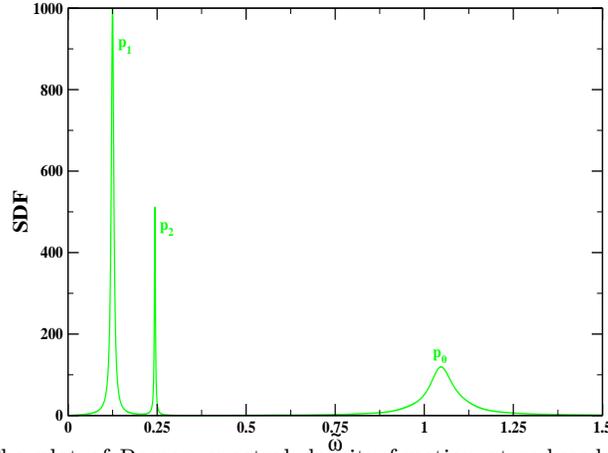
The plot of Raman spectral density function at reduced temperature $t=0.001~(z=0.00952,~{e^{\prime}}=0.0136)$ for the above set of values and the suitable values of bare phono frequency $p=0.08$ and spectral width $e_1=0.004$.}
%%%%%%%%%%%%%%%%%%%%%%%%%%%%%%%%%%%%%%%%%%%%%%%%%%%%%%%%%%%%%%%%%%%%%%%%%%%%%%%

Figure 3 shows the plot of SDW vs $\tilde{\omega}$ for different values
 of temperature $t=0-0.0023$ and for other fixed parameters. The 
self-consistent plot of $z$ and $e^{\prime}$ shown in figure 1 indicates
 that the SC order and JT energy coexist upto the SC transition 
temperature $t_c\simeq 0.0052$. The temperature dependence of SDF
 displays the interplay of the SC and JT gaps as shown in figure 3.
 With the increase of temperature from $t=0$ to $0.0023$ in the 
coexistence phase the peak position of the SC gap move towards higher
 energies and the peak position of the JT gap excitation peak moves towards
 lower energies and merge each other at temperature $t=0.0022$. On further
 increasing the temperature, the spectral height of the merged peak is 
suppressed accompanied by a larger spectral width and finally vanish at
 $t>0.0023$, which is smaller than the SC transition temperature. This
 SC peak in coexistence phase vanishes due to the interaction of static
 and dynamic JT effect displaying a strong interaction between them.

\vspace{0.2in}

%%%%%%%%%%%%%%%%%%%%%%%%%%%%%%%%%%%%%%%%%%%%%%%%%%%%%%%%%%%%%%%%%%%%%%%%%%%%%
\bigskip

{\vbox{
\begin{center}
      \epsfig{file=rmn1.eps,width=8cm,height=6cm}
      \vspace{-1.0cm}
 %     \begin{figure}[p]
  %    \label{c1f1}
   %   \vspace{0.5cm}
    %  \end{figure}
\end{center}
}}

\vspace{0.5cm}

{\small {\bf Fig.} \ {\bf 3} \
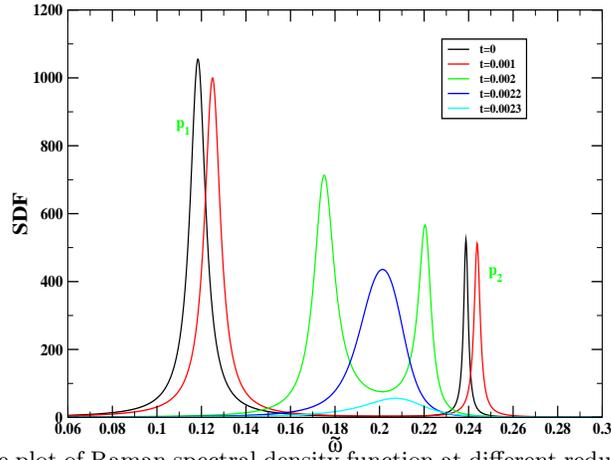
The plot of Raman spectral density function at different reduced temperatures $t=0.0,~0.001,~0.002,~0.0022$ and $0.0023$.}
%%%%%%%%%%%%%%%%%%%%%%%%%%%%%%%%%%%%%%%%%%%%%%%%%%%%%%%%%%%%%%%%%%%%%%%%%%%%%%%

\vspace{0.2in}

%%%%%%%%%%%%%%%%%%%%%%%%%%%%%%%%%%%%%%%%%%%%%%%%%%%%%%%%%%%%%%%%%%%%%%%%%%%%%
\bigskip

{\vbox{
\begin{center}
      \epsfig{file=rmn2.eps,width=8cm,height=6cm}
      \vspace{-1.0cm}
 %     \begin{figure}[p]
  %    \label{c1f1}
   %   \vspace{0.5cm}
    %  \end{figure}
\end{center}
}}

\vspace{0.5cm}

{\small {\bf Fig.} \ {\bf 4} \
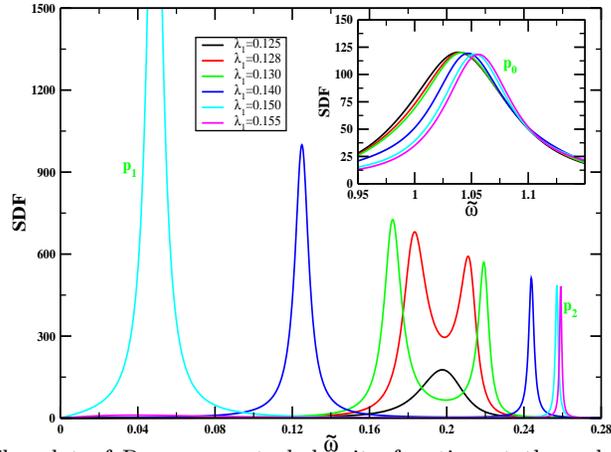
The plot of Raman spectral density function at the reduced temperature $t=0.001$ for different values of the normal EP coupling $\lambda_1=0.125,~0.128,~0.13,~0.0,~0.15$ and $0.155$.}
%%%%%%%%%%%%%%%%%%%%%%%%%%%%%%%%%%%%%%%%%%%%%%%%%%%%%%%%%%%%%%%%%%%%%%%%%%%%%%%

Figure 4 shows the effect of normal EP coupling $(\lambda_1)$ on the two 
Raman active peaks. At very low value of EP coupling $(\lambda_1=0.125)$ 
the two excitation peaks merge with each other and appear as one peak at 
energy $\tilde{\omega}=0.2$. As $\lambda_1$ increases, the single peak 
splits into two separating the low energy SC excitation peak $p_1$ from the
 high energy JT distortion peak $p_2$. With the further increase of 
$\lambda_1$ the JT peak $p_2$ moves to higher energies and the SC peak 
$p_1$ move to lower energies, finally for a higher value of normal EP 
coupling $\lambda_1\simeq 0.155$, the SC peak $p_1$ nearly vanishes and
 the high energy JT peak exists. The inset of figure 4 shows the effect of the
 normal EP coupling on the bare phonon frequency peak $p_0$. When the normal
 EP coupling $\lambda_1$ increases from $0.125$ to $0.155$,
 the spectral weight of
 bare phonon frequency $p_0$ shifts to higher energies exhibiting the
 hardening behaviour of the phonon mode.

Figure 5 shows the SDF vs $\tilde{\omega}$ for different values of
 DJT coupling $(\lambda_2)$. In absence of DJT coupling 
$(\lambda_2=0)$ two peaks $p_1$ and $p_2$ appear due to the normal
 EP coupling $(\lambda_1)$. With increase of DJT coupling the JT 
excitation peak $p_2$ shifts towards the lower energies with slightly
 enhancement in spectral height. This decrease in JT energy of the 
peak $p_2$ with increase of DJT coupling is consistent with our 
earlier result obtained from the self-consistent solution of SC gap 
and JT energy gap\cite{Raj}. On the other hand with increase of the 
DJT coupling $(\lambda_2)$ the position of the SC excitation peak shifts
 to higher energies with a monotonically decrease in spectral height. 
This suggests that the DJT coupling suppresses the JT gap and consequently
 enhances the SC gap. The inset of figure 5 shows the effect of the
 DJT coupling on the bare phonon frequency peak $p_0$. When the DJT coupling
 is increased from $\lambda_2=0.0$ to $0.25$, the bare phonon frequency shows
 neither appreciable change in its spectral width nor its spectral
 height.

\vspace{0.2in}

%%%%%%%%%%%%%%%%%%%%%%%%%%%%%%%%%%%%%%%%%%%%%%%%%%%%%%%%%%%%%%%%%%%%%%%%%%%%%
\bigskip

{\vbox{
\begin{center}
      \epsfig{file=rmn3.eps,width=8cm,height=6cm}
      \vspace{-1.0cm}
 %     \begin{figure}[p]
  %    \label{c1f1}
   %   \vspace{0.5cm}
    %  \end{figure}
\end{center}
}}

\vspace{0.5cm}

{\small {\bf Fig.} \ {\bf 5} \
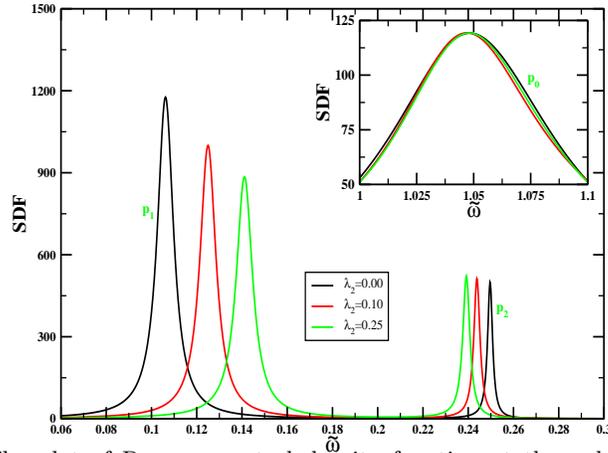
The plot of Raman spectral density function at the reduced temperature $t=0.001$ for different values of the dynamic EP coupling $\lambda_2=0.0,~0.10$ and $0.25$.}
%%%%%%%%%%%%%%%%%%%%%%%%%%%%%%%%%%%%%%%%%%%%%%%%%%%%%%%%%%%%%%%%%%%%%%%%%%%%%%%

Figure 6 shows the plot SDF vs $\tilde{\omega}$ for different values of 
SC coupling from $g=0.0280$ to $0.0350$. For a moderately high value of 
SC coupling $g=0.0350$, the SC excitation peak $p_1$ and the JT gap 
excitation peak $p_2$ are well separated and the JT excitation peak 
$p_2$ shifts to lower energies with gradual increase of spectral height.
 This suggests that with decrease of SC coupling the JT gap is suppressed 
and in consequence the SC gap is enhanced considerably. On further 
decreasing SC coupling, two excitation peaks appear to merge for 
$g=0.0285$ and finally merged for $g=0.0280$ with considerably suppression 
of the spectral height and with more decrease of the SC coupling the 
merged peak will vanish completely. The inset of figure 6 shows the effect
 of the SC coupling on the bare phonon frequency peak $ p_0$. When the SC
 coupling increases, the spectral weight shifts from lower to the higher
 energies exhibiting the hardening behaviour of the phonon mode and the
 spectral height shows no appreciable change.

\vspace{0.2in}

%%%%%%%%%%%%%%%%%%%%%%%%%%%%%%%%%%%%%%%%%%%%%%%%%%%%%%%%%%%%%%%%%%%%%%%%%%%%%
\bigskip

{\vbox{
\begin{center}
      \epsfig{file=rmn4.eps,width=8cm,height=6cm}
      \vspace{-1.0cm}
 %     \begin{figure}[p]
  %    \label{c1f1}
   %   \vspace{0.5cm}
    %  \end{figure}
\end{center}
}}

\vspace{0.5cm}

{\small {\bf Fig.} \ {\bf 6} \
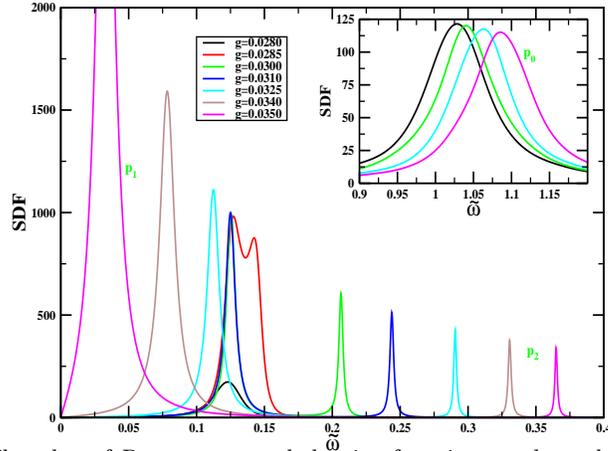
The plot of Raman spectral density function at the reduced temperature
$t=0.001$ for different values of the SC coupling
$g=0.0280,~0.0285,~0.0300,~0.0310,0.0325,~0.0340$ and $0.0350$.}
%%%%%%%%%%%%%%%%%%%%%%%%%%%%%%%%%%%%%%%%%%%%%%%%%%%%%%%%%%%%%%%%%%%%%%%%%%%%%%%

\vspace{0.2in}

%%%%%%%%%%%%%%%%%%%%%%%%%%%%%%%%%%%%%%%%%%%%%%%%%%%%%%%%%%%%%%%%%%%%%%%%%%%%%
\bigskip

{\vbox{
\begin{center}
      \epsfig{file=rmn5.eps,width=8cm,height=6cm}
      \vspace{-1.0cm}
 %     \begin{figure}[p]
  %    \label{c1f1}
   %   \vspace{0.5cm}
    %  \end{figure}
\end{center}
}}

\vspace{0.5cm}

{\small {\bf Fig.} \ {\bf 7} \
The plot of Raman spectral density function at the reduced temperature $t=0.001$ for different values of the JT coupling constant $g_1=0.146,~0.147,~0.149,~0.152,~0.155,~0.157$ and $0.159$.}
%%%%%%%%%%%%%%%%%%%%%%%%%%%%%%%%%%%%%%%%%%%%%%%%%%%%%%%%%%%%%%%%%%%%%%%%%%%%%%%

Figure 7 shows the effect of static JT coupling on Raman spectral
 peaks of the system. For a lower static JT coupling $g_1=0.146$, 
the JT excitation peak $p_2$ appears at higher energy
 $\tilde{\omega}=\simeq 0.265$
 but the SC excitation peak appears as a very flat peak at lower energy. 
With increase of JT coupling, the JT excitation peak $p_2$ shifts towards
 the lower energies indicating the suppression of JT gap while the SC 
excitation peak position shifts towards higher energies with a 
corresponding decrease in spectral height, indicating the enhancement
 of SC gap with $g_1$. This is consistent with our conclusion from
 figure 1 that the suppression of JT energy induces the enhancement
 of magnitude of the SC gap. Finally for a static JT coupling $g_1=0.159$,
 two excitation peaks tend to merge to give rise a single peak, 
where the SC gap tends to be equal to the JT gap. With the further
 increase of the JT coupling the two peaks will merge with decrease
 in spectral height and will finally be suppressed completely. The inset of
 figure 7 shows the effect of the static JT coupling on the bare phonon
 frequency peak $p_0$. It is observed that, with the increase of the static JT
 coupling $g_1$, the spectral weight of the phonon frequency mode shifts to
 lower energies exhibiting its softening behaviour, but there is no
 appreciable change in spectral height but some increase in spectral width.

\vspace{0.2in}

%%%%%%%%%%%%%%%%%%%%%%%%%%%%%%%%%%%%%%%%%%%%%%%%%%%%%%%%%%%%%%%%%%%%%%%%%%%%%
\bigskip

{\vbox{
\begin{center}
      \epsfig{file=rmn6.eps,width=8cm,height=6cm}
      \vspace{-1.0cm}
 %     \begin{figure}[p]
  %    \label{c1f1}
   %   \vspace{0.5cm}
    %  \end{figure}
\end{center}
}}

\vspace{0.5cm}

{\small {\bf Fig.} \ {\bf 8} \
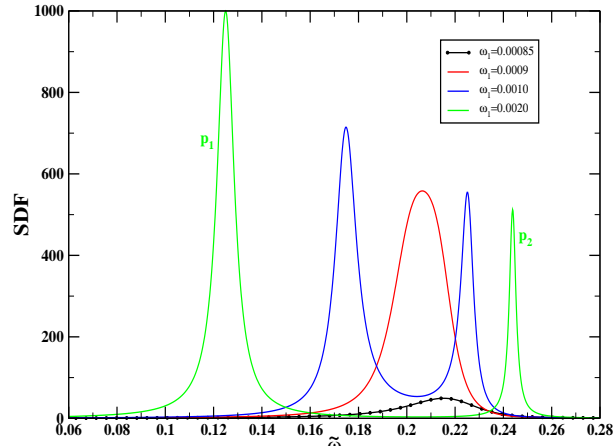
The plot of Raman spectral density function at the reduced temperature $t=0.001$ for different values of the phonon vibrational frequency $\omega_1=0.00085,~0.0009,~0.0010$ and $0.0020$.}
%%%%%%%%%%%%%%%%%%%%%%%%%%%%%%%%%%%%%%%%%%%%%%%%%%%%%%%%%%%%%%%%%%%%%%%%%%%%%%%

Figure 8 shows the plot of SDF vs $\tilde{\omega}$ for different
 values of phonon vibrational frequency $\omega_1$ at any finite 
temperature. It is observed from the figure that a single peak appears
 in the SDF for very low phonon vibrational frequency $\omega_1=0.00085$.
 The spectral height of the single peak is enhanced with increase of 
phonon vibrational frequency to $\omega_1=0.0009$. On further increase
 in $\omega_1$ to $0.0010$ and $0.0020$ the single peak splits into two 
giving rise to high frequency JT excitation peak $p_2$ and
 SC gap excitation peak $p_1$ at low frequency. 
In other hand the increase of frequency $\omega_1$ the magnitude of the 
JT gap is enhanced to higher energy, while the SC gap is suppressed to a
 lower energy. The separation between the two gaps gradually increases 
with the increase of $\omega_1$ resulting in weak interplay between these
 two interactions.

\vspace{0.2in}

%%%%%%%%%%%%%%%%%%%%%%%%%%%%%%%%%%%%%%%%%%%%%%%%%%%%%%%%%%%%%%%%%%%%%%%%%%%%%
\bigskip

{\vbox{
\begin{center}
      \epsfig{file=rmn7.eps,width=8cm,height=6cm}
      \vspace{-1.0cm}
 %     \begin{figure}[p]
  %    \label{c1f1}
   %   \vspace{0.5cm}
    %  \end{figure}
\end{center}
}}

\vspace{0.5cm}

{\small {\bf Fig.} \ {\bf 9} \
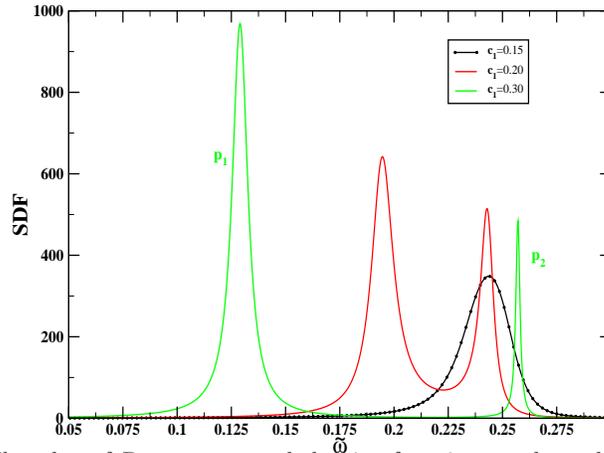
The plot of Raman spectral density function at the reduced temperature $t=0.001$ for different values of the reduced incident photon frequency $c_1=0.15,~0.20$ and $0.30$.}
%%%%%%%%%%%%%%%%%%%%%%%%%%%%%%%%%%%%%%%%%%%%%%%%%%%%%%%%%%%%%%%%%%%%%%%%%%%%%%%

Figure 9  shows the plot of SDF vs $\tilde{\omega}$ for different
 values of incident photon frequency $c_1=\frac{\omega}{2t_0}$. 
For lower incident photon frequency $c_1=0.15$ a single symmetric
 excitation peak appears at energy $\tilde{\omega}\simeq 0.24$ in the SDF.
 Here the SC gap energy and JT gap energy are equal. With increase of
 photon frequency $c_1$ to $0.20$ and $0.30$ the single peak splits 
into two giving rise to the low frequency SC gap excitation peak $p_1$ 
at lower energies and the high frequency JT gap excitation
 peak $p_2$ at higher energies.
 Thus this is consistent with our conclusion from figure 1 that the 
increase of JT distortion gap tends to suppress the SC gap and this shows
 a clear interplay of SC and JT order parameters.
%\newpage
\section{Conclusion}
 The present calculation has been carried out for a system which 
exhibits the coexistence of $s$-wave type superconducting state and the static
 and dynamic Jahn-Teller distortion. The results of this calculation
 may be applied to some cuprate superconductors like the bismuthates\cite{rf7a}.
 The phonon SDF is calculated by the Green's 
function technique for the coexistence phase of the superconductivity and
 lattice strain in presence of DJT coupling. The SDF displays three peaks 
i.e., the peak $p_0$ centered at energy $\tilde{\omega}\simeq 1.05$ for the
 renormalized bare phonons, the peak $p_1$ at energy 
$\tilde{\omega}\simeq 0.125$  representing the SC gap excitation and
 the peak $p_2$ at energy $\tilde{\omega}\simeq 0.245$ representing 
the gap excitation associated with the JT distortion. The interplay of
 these two long range orders are investigated by varying the normal EP
 coupling ($\lambda_1$), the static JT coupling ($g_1$), the DJT coupling 
($\lambda_2$), the SC coupling ($g$), temperature ($t$), phonon vibrational
 frequency ($\omega_1$) and incident photon frequency ($c_1$). It is observed
 that the normal EP coupling ($\lambda_1$) enhances the JT gap and suppresses
 the SC gap. As a result JT excitation peak shifts to the higher energies 
while SC gap excitation peak shifts to the lower energies. It is observed 
that both the static and dynamic Jahn-Teller couplings $g_1$ and $\lambda_2$
 suppress the insulating JT gap and enhance the SC gap and as a result with 
increase of $g_1$ and $\lambda_2$ the JT excitation peak shifts to the lower 
energies and the position of the SC excitation peak shifts to the higher
 energies. From the study of the effects of phonon vibrational frequency
 ($\omega_1$) and the incident photon frequency ($c_1$) it is observed that
 both the frequencies $\omega_1$ and $c_1$ induce higher JT distortion 
leading to suppression of the SC gap. Under this condition of increasing
 $\omega_1$ and $c_1$ the SC excitation peak $p_1$ and JT excitation peak
 $p_2$ are gradually separated from each other. The present study displays
 a very strong interplay between SC interaction, static JT interaction as
 well as DJT interaction.

\section*{Acknowledgements}
The authors gracefully acknowledge the research facilities offered
 by the Institute of Physics, Bhubaneswar, India during their short stay.
%\newpage

\end{document}